\documentclass[Conference] {IEEEtran}
\IEEEoverridecommandlockouts
\usepackage{cite}
\usepackage[utf8]{inputenc}
\usepackage{amsmath,amssymb,amsfonts}
\usepackage{algorithmic}
\usepackage{graphicx}
\usepackage{textcomp}
\usepackage{longtable}
\usepackage{array}
\usepackage{xcolor}
\usepackage{color}
\usepackage{arydshln}
\usepackage{blindtext, graphicx}
\usepackage{listings}
\usepackage{multirow}
\usepackage{graphicx}
\usepackage[font=small]{caption}
\usepackage[utf8]{inputenc}
\def\BibTeX{{\rm B\kern-.05em{\sc i\kern-.025em b}\kern-.08em
    T\kern-.1667em\lower.7ex\hbox{E}\kern-.125emX}}

\begin{document}
\title{Towards Determining the Effect of Age and Educational Level on Cyber-Hygiene}

\author{
\IEEEauthorblockN{Celestine Ugwu\textsuperscript{1}, Casimir Ani  \textsuperscript{2}, Modesta Ezema \textsuperscript{1}, Caroline Asogwa \textsuperscript{1}, Uchenna Ome\textsuperscript{1}, \\Adaora Obayi\textsuperscript{1}, Deborah Ebem\textsuperscript{1}, Aminat Atanda \textsuperscript{1}, Elochukwu Ukwandu \textsuperscript{3}}

\IEEEauthorblockA{\textsuperscript{1}Dept. of Computer Science.  Faculty of Physical Science, University of Nigeria, Nsukka,
 Enugu State, Nigeria.\\ \textsuperscript{2}Dept. of Philosophy.  Faculty of Social Science, University of Nigeria, Nsukka, Enugu State, Nigeria.\\ \textsuperscript{3}Dept. of Computer Science. Cardiff School of Technologies, Cardiff Metropolitan University, United Kingdom.}\\
E-mail addresses: {(celestine.ugwu, casmir.ani, modesta.ezema, caroline.asogwa, uchenna.ome,\\adaora.obayi, debora.ebem, aminat.)@unn.edu.ng}, eaukwandu@cardiffmet.ac.uk\\
Corresponding author: Elochukwu Ukwandu
}

\maketitle
\begin{abstract}
 As internet related challenges increase such as cyber-attacks, the need for safe practises among users to maintain computer system's health and online security have become imperative, and this is known as cyber-hygiene. Poor cyber-hygiene among internet users is a very critical issue undermining the general acceptance and adoption of internet technology. It has become a global issue and concern in this digital era when virtually all business transactions, learning, communication and many other activities are performed online. Virus attack, poor authentication technique, improper file backups and the use of different social engineering approaches by cyber-attackers to deceive internet users into divulging their confidential information with the intention to attack them have serious negative implications on the industries and organisations, including educational institutions. Moreover, risks associated with these ugly phenomena are likely to be more in developing countries such as Nigeria. Thus, authors of this paper undertook an online pilot study among students and employees of University of Nigeria, Nsukka and a total of 145 responses were received and used for the study. The survey seeks to find out the effect of age and level of education on the cyber hygiene knowledge and behaviour of the respondents, and in addition, the type of devices used and activities they engage in while on the internet. The result of the study revealed that the independent variables, age and level of education do not have statistical significance on the cyber hygiene knowledge and behaviour of the respondents. It was equally shown from the result of the survey that mobile phone is the most widely used device with 93.1\%, followed by laptop with 71\%. The activity that these respondents perform mostly while online is internet browsing with 94.5\%, followed by online learning with 86.9\% according to our research. From our findings also, we discovered that only 53.79\% of the respondents have good cyber hygiene knowledge, while 51.72\% have good cyber hygiene behaviour. Our findings show wide adoption of internet in institution of higher learning, whereas, significant number of the internet users do not have good cyber hygiene knowledge and behaviour. Hence, our findings can instigate an organised training for students and employees of higher institutions in Nigeria.


\end{abstract}

\section{\textbf{Introduction}}
\label{Section: Introduction}
The wide penetration and overwhelming acceptance of internet in Nigeria and across the globe has facilitated learning, business and other activities that are internet enabled. The introduction of internet technology has impacted positively on everyday activities in the various sectors of human endeavours. However, this technological advancement has equally resulted to increased cyber threats, vulnerabilities and risks. In Nigeria, for instance, internet has made it possible the perpetration of different forms of cyber-crime on daily basis ranging from fraudulent electronic mails, pornography, identity theft, hacking, cyber harassment, spamming, Automated Teller Machine spoofing, piracy and phishing~\cite{omodunbi2016cybercrimes}. \\ 

According to~\cite{adesina2017cybercrime}, in the year 2016, the personal information of  millions of people were stolen through cyber-crime, which comprises of 40 million people in United States of America (USA), 54 million in Turkey, 20 million in Korea, 16 million in Germany and over 20 million in China.\\ 

Moreover, these risks associated with this ugly phenomenon are likely to be more in developing countries such as Nigeria. In Nigeria, cyber-attacks are committed by people of different age ranges, both old and young are involved in this act though majority are young ones~\cite{hassan2012cybercrime}. The cyber space is the main channel through which financial fraud is being perpetrated in the Nigerian banking industry~\cite{ibrahimimpact}. This therefore calls for concerted effort to curb the menace of cyber-crime activities on individuals, organisations and governments. \\

One of the ways through which cyber-crime can be mitigated is by improving the cyber-hygiene culture of the internet users. Cyber-hygiene here refers to those cyber-security attitudes and behaviours which internet users are expected to adopt to ensure the safety and integrity of their data and also their devices in the case of cyber-attacks by the internet fraudsters~\cite{vishwanath2020cyber}. \\

This research through a pilot study seeks to find out if there is relationship between demographic factors (age and level of education) and cyber-hygiene among students and employees of University of Nigeria, Nsukka. An ethical approval was obtained from the university for the purpose of this research. This paper presents the result of the survey for determining the relationship between age and level of education of students and staff, and cyber-hygiene. The wide adoption of internet in tertiary institutions, coupled with quest for migration to online learning as a result o f the outbreak of Coronavirus-2019 (COVID-19) necessitated this research.\\
The remaining parts of this paper are organised as follows:
Section~\ref{Section: Review} presents the literature review, followed by methodology as Section~\ref{Section: Method}. Section~\ref{Section: Discussion} presents the Result and Discussion of the research analysis. The Conclusion forms the final section of the the paper. \\

\section\textbf{{Review of Related Literature}}
\label{Section: Review}
While there have been several studies involving various users and aspects of cyber-hygiene, there is currently is no substantial survey, which explores cyber-hygiene by considering individual differences and the users’ level of knowledge especially amongst students and employees in the universities in Nigeria.  The main reasons for this research are to find out the effect of age and educational level on cyber-hygiene and also to ascertain the level of cyber-hygiene knowledge and behaviour among students and employees of University of Nigeria, Nsukka.\\

 Previously many researchers have made findings on cyber-hygiene cultures of internet users. Talib~\textit{et al.}~\cite{talib2010analysis} found that 97\% of users did have antivirus software at home. Authors in~\cite{talib2010analysis} also reported that 72\% of people who are not trained on the topic did use firewall protection. There are also discrepancies in the data that describes the use of Spam protection. \\
 
The study conducted in~\cite{neigel2020holistic} focused on understanding the human factors and individual differences that influence cyber-hygiene. The results demonstrated in the work indicate that cyber-hygiene education need not target a particular sex or age group in terms of content or delivery method, which contradicts previous findings from~\cite{whitty2015individual}.
The research carried out by authors in~\cite{whitty2015individual} highlighted the importance of understanding the types of people who are more likely to engage in the risky behavior of sharing passwords. They found a number of variables that can be used to predict users that engage in the risky practice of sharing passwords, which are age, perseverance, and self-monitoring.\\

Research conducted in~\cite{barrerainfluence} seeks to examine relationships, if any, between cyber-security awareness level and the background of participants pursuing careers in the area of Information Systems (IS) and/or Information Technology (IT) at the bachelor’s level in three different geographic locations; Germany, United Kingdom, and the United States of America, focusing specifically on four demographic variables: gender, age, education Level Completed and Current Employment Status. They arrived at a conclusion that Awareness is frequently associated to operational situations, where specific reasons require individuals to have an identifiable awareness level for a specific context. Therefore, individuals and business organisations benefit from higher levels of security awareness, which ultimately reflects higher literacy levels and learning. Lastly, business continuity depends on how individuals respond to various situations, exercise caution in their decisions, and ultimately, how aware they are about current and future security risks in their doings.\\

A study to create awareness of security threat and avoidance was carried out in \cite{arachchilage2014security}. The study was carried out on anti-phishing education to guard against identity theft and related issues. It examined whether the knowledge of cyber-hygiene concepts has effect on users’ ability to avoid phishing attack. An online questionnaire were distributed and collected from 161 computer users from Brunel University and the University of Bedfordshire and their responses were used for analysis. The finding revealed that the knowledge of concepts enables internet users to avoid phishing attacks. It was concluded that educating users on the knowledge of concepts has significant positive effects on enabling users to avert phishing threats and attacks.\\

\section{Methodology}
\label{Section: Method}
\subsection{\textbf{Research Goal}}
The primary goal of this research is to determine the significance effect of certain demographic factors/variables such as age and level of education on cyber-hygiene culture among students and employees of University of Nigeria, Nsukka. The reasons for choosing the university are for convenience, cost reduction and availability of participants to researchers. In respect to this research, an ethical approval was sought from the university authority and it was given. Because of prolonged Academic Staff Union of Universities in Nigeria and lockdown necessitated by COVID-19 pandemic, the response rate was very low. Based on this reason, the researchers decided to take a pilot study from the received responses to test the respondent’s understanding of the research questionnaire. An updated version of the survey will be conducted as soon as the university reopens.\\

\subsection{\textbf{Instrument and Method of Data Collection}}
Questionnaire was the basic instrument used to gather required data from the chosen institution.  A well-structured questionnaire was designed with Google form for the survey and distributed via online approach using the institution’s online mailing system and group WhatsApps to the chosen participants. Responses from the respondents were collated in a datasheet and were subjected for analysis.\\

\subsection{\textbf{Sampling Technique}}
Convenience sampling, a common non-probability sampling was applied in selecting the sample used for the survey. This sampling method is fast, cost-effective and it makes sample easily available.\\

\subsection{\textbf{Sample Size}}
A total of 145 responses were received and used for this pilot study. The respondents cut across different age ranges and educational levels to accommodate all relevant demographics.\\

\subsection{\textbf{Research Model}}
Here in our study, the analysis of the effect of some selected demographic factors on cyber-hygiene among students and employees of University of Nigeria, Nsukka was carried out. The demographic factors represent the independent variables, which include age and educational level whereas cyber-hygiene represents the dependent variable in this study. The percentage of internet usage for educational purposes increases with the student’s age, which shows the increasing e-learning prospect with the level of education as observed in \cite{tirumala2016survey}. Some common aspects of cyber-hygiene culture were chosen for this study namely, storage and virus attack hygiene, social network hygiene, authentication hygiene and social engineering hygiene. The first three which include virus attack hygiene, social network hygiene and authentication hygiene were adopted from~\cite{vishwanath2020cyber} with little modification in nomenclature.  Social engineering hygiene was introduced by the research team as users’ cyber-hygiene dimension required to overcome the cyber-security threat considered in~\cite{fatokun2019impact}. Questions in section B and section C of the questionnaire were spread across these four aspects of cyber-hygiene. The association of the independent variables and the dependent variable is shown in Figure~\ref{fig:my_label} below.\\

\begin{figure} [h]
    \centering
    \includegraphics[width=8cm,height=4cm]{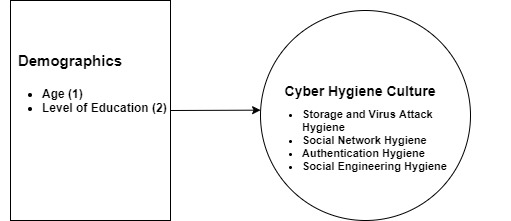}
    \caption{\textbf{Research Model}}
    \label{fig:my_label}
\end{figure}

\subsection{\textbf{Research Questions}}
In order to ascertain if there is any significance relationship between demographic factors and cyber-hygiene culture, the following research questions were raised:
\begin{enumerate}
    \item Does age of internet users have any significant relationship with cyber-hygiene culture?
    \item Does level of education of internet users have any significant relationship with cyber-hygiene culture?
\end{enumerate}

\subsection{\textbf{Research Hypotheses}}
Two null hypotheses were formulated as follows by the researchers to answer the above research questions:
\begin{itemize}
    \item H1: Age of internet user does not have any significant effect on cyber-hygiene
    \item H2: Level of education of internet user does not have any significant effect on cyber-hygiene
\end{itemize}

\subsection{\textbf{Data Analysis}}
Our research questionnaire in this study consist of three sections, section A which was used to gather information on demographic factors, type of activities they perform with internet and the type of devices used for these activities. Questions on section B were asked to determine participants’ knowledge on threat and concept. This section has total of 12 questions, which comprises of multiple choice questions and five points Likert scale statements. Options for the five points Likert scale statements ranging from \textbf{Strongly Agree}, \textbf{Agree}, \textbf{Don’t Know}, \textbf{Disagree} and \textbf{Strongly Disagree}. In each of the statement, either Strongly Agree and Agree or Disagree and Strongly Disagree are considered as the correct option. Section C seeks to verify the level of cyber-hygiene culture exhibited by participants or their cyber-hygiene behaviour. On section C, respondents were asked to report the degree to which they engage on certain cyber-hygiene behaviour on a four points Likert scale with option ranging from Every time, Often, Rarely,  and Never. In this case, either Every time and Often or Never is considered as the correct option. Questions in section B and C concentrates within the four chosen aspects of cyber-hygiene (\textbf{storage and virus, social network, authentication and social engineering}) used for this study. The respondents were asked to go through the definition of some technical terms such as cyber-hygiene, virus, authentication, social network and social engineering used in the study for clarification, easy understanding and better response. \\

According to~\cite{akman2010gender}, the test of statistical hypotheses is one of the main areas of statistical inference. Version 20 of the Statistical Packages for Social Sciences (SPSS) software was used for the analysis. Chi-square test, an analytical tool embedded in Statistical Packages for Social Sciences (SPSS) software was used to test the significance of these demographic variables. Also, multiple regression technique was subsequently used to determine the direction of the main impacts of the independent variables on the dependent one. Out of these tests, the following inferences made on this paper were drawn.\\

\section{\textbf{Result and Discussion}}
\label{Section: Discussion}
This research survey sought to find out from the participants the type of devices they use for internet, the activities they use internet for, their knowledge of cyber-hygiene concepts and threats, and  their cyber-hygiene behaviour. Our findings are shown as tables and discussed in the following sections below:\\

\subsection{\textbf{Devices Used for the Internet}}
As shown in Table~\ref{tab: table1}, it was found that majority of the respondents, 135(93.1\%) indicated mobile phone as the most commonly used device, followed by laptop 103(71\%). The least used is other devices 2(1.4\%), followed by desktop 17(11.7\%) and finally, Tablet with 21(14.5\%). This varies from the result in~\cite{cain2018exploratory}, where the number of laptop users are more than the number of smart phone users. This may not be far from the low cost of mobile phones and its proliferation across the country compared to laptops.\\

\begin {table}[h!]
  \begin {center}
    \caption{\textbf{Devices used for internet}}
      \label{tab: table1}
       \begin{tabular}{l|c|r}
         \textbf{Variable} & \textbf{Frequency} & \textbf{Percentage}\\
         \hline
          Laptop & 103 & 71\\
          Desktop & 17 & 11.7\\
          Mobile phone & 135 & 93.1\\
         Tablet & 21 & 14.5\\
        Other devices & 2 & 1.4\\
      \end{tabular}
  \end {center}
\end {table}

\subsection{\textbf{Uses of Internet}}
Table~\ref{tab: table2} shows various purposes for which the participants use internet for. It can be drawn from the table that apart from playing games with only 23.4\% (34, n=145), other purposes in which the participants use the internet for attracts high percentage of users ranging from 50.3 to 94.5. The highest is internet browsing with 94.5\% (137, n=145), followed by learning with 86.9\% (126, n=145). This findings show high level of adoption and usage of internet among students and employees of higher institutions, which calls for survey of their cyber-hygiene knowledge and behaviour.\\

\begin {table}[h!]
  \begin {center}
    \caption{\textbf{Uses of internet}}
      \label{tab: table2}
       \begin{tabular}{l|c|r}
         \textbf{Variable} & \textbf{Frequency} & \textbf{Percentage}\\
         \hline
          Browsing & 137 & 94.5\\
          Email & 117 & 80.7\\
          Downloading music/video & 96 & 66.2\\
         Social Networking & 118 & 81.4\\
        Playing games & 34 & 23.4\\
        Learning & 126 & 86.9\\
        Business & 73 & 50.3\\
        Banking & 91 & 62.8\\
      \end{tabular}
  \end {center}
\end {table}

\subsection\textbf{{Demographics of Respondents}}
 From the demographic information presented in Table~\ref{tab: table3}, the age of the respondents were coded into five groups, respondents between 15-24years, 25-34year, 35-44years, 45-54years and above 54years. The distributions of respondents within these groups are as follows: 15-24years 40.7\% (59, n=145), 25-34year 25.5\% (37, n=145), 35-44years 20.0\% (29, n=145), 45-54years 11.0\% (16, n=145) and above 54years 2.8\% (4, n=145). As regards their educational level of the respondents, 0.7\% (1) has diploma, 44.1\% (64) are undergraduates, 21.4\% (31) are graduates, 20.0\% (29) are Masters degree holders and 13.8\% (20) have Doctorate degree.\\
 
 \begin {table}[h!]
  \begin {center}
    \caption{\textbf{Socio-demographics of the respondents}}
      \label{tab: table3}
       \begin{tabular}{l|c|r}
         \textbf{Variable} & \textbf{Frequency} & \textbf{Percentage}\\
         \hline
          \multirow {2} {*}
            Age\\
          15-24yrs & 59 & 40.7\\
          25-34yrs & 37 & 25.5\\
          34-44yrs & 29 & 20.0\\
         45-54yrs & 16 & 11.0\\
        Above 54yrs & 4 & 2.8\\
        \hline
        Educational Level\\
        Diploma & 1 & 0.7\\
        Undergraduate & 64 & 44.1\\
        Graduate & 31 & 21.4\\
        Masters & 29 & 20.0\\
        Doctorate & 20 & 13.8\\
      \end{tabular}
  \end {center}
\end {table}

\subsection{\textbf{Age and Cyber hygiene}}
In order to find out the effect of the age of the participants on cyber-hygiene, chi-square tests were carried out to check the relationship between age and cyber-hygiene knowledge and also between age and cyber-hygiene behaviour. From the result of the chi-square test to determine the relationship between age and cyber-hygiene as shown in Table~\ref{tab: table4}, the p-value of (0.455), which is greater than 0.05 indicates no statistical significance. This means that there is no relationship between age and cyber-hygiene knowledge.
On the same hand, Table~\ref{tab: table5} presents a p-value of 0.551 for age and cyber-hygiene behaviour, meaning that age of the participants has no relationship with the cyber-hygiene behaviour. The result here conforms to what was obtained in-\cite{cain2018exploratory}.\\

\begin {table}[h!]
  \begin {center}
    \caption{\textbf{Chi-Square Test for Age and Cyber hygiene knowledge}}
      \label{tab: table4}
       \begin{tabular}{l|c|c|r}
        \textbf{Quantity} & \textbf{Value} & \textbf{df} & \textbf{ Asym. Sig (2-sided)}\\
         \hline
          Pearson Chi-Square & 3.691 & 4 & 0.455\\
          Likelihood Ratio & 3.175 & 4 & 0.270\\
          Linear-by-Linear Association & 0.896 & 1 & 0.344\\
         N of Valid Cases & 145 \\
      \end{tabular}
  \end {center}
\end {table}

\begin {table}[h!]
  \begin {center}
    \caption{\textbf{Chi-Square Test for Age and Cyber hygiene behaviour}}
      \label{tab: table5}
       \begin{tabular}{l|c|c|r}
        \textbf{Quantity} & \textbf{Value} & \textbf{df} & \textbf{ Asym. Sig (2-sided)}\\
         \hline
          Pearson Chi-Square & 3.043 & 4 & 0.551\\
          Likelihood Ratio & 3.093 & 4 & 0.542\\
          Linear-by-Linear Association & 0.301 & 1 & 0.583\\
         N of Valid Cases & 145 \\
      \end{tabular}
  \end {center}
\end {table}

\subsection{\textbf{Level of Education and Cyber hygiene}}
Chi-square tests were also conducted in two categories to determine the relationship between level of education as an independent variable and the dependent variable, cyber-hygiene. In the first category, the p-value of the chi-square test for level of education and cyber-hygiene knowledge as shown in Table \ref{tab: table6} is 0.628. The interpretation is that the level of education of the respondent does not have relationship with cyber-hygiene knowledge. Secondly, the chi-square test for level of education and cyber-hygiene behaviour has p-value of 0.285 as shown in Table \ref{tab: table7}. This also means that there is no relationship between the level of education of the respondents and cyber-hygiene behaviour.\\

\begin {table}[h!]
  \begin {center}
    \caption{\textbf{Chi-Square Test for Level of Education and Cyber-hygiene knowledge}}
      \label{tab: table6}
       \begin{tabular}{l|c|c|r}
        \textbf{Quantity} & \textbf{Value} & \textbf{df} & \textbf{ Asym. Sig (2-sided)}\\
         \hline
          Pearson Chi-Square & 2.596 & 4 & 0.628\\
          Likelihood Ratio & 2.985 & 4 & 0.560\\
          Linear-by-Linear Association & 0.655 & 1 & 0.418\\
         N of Valid Cases & 145 \\
      \end{tabular}
  \end {center}
\end {table}

\begin {table}[h!]
  \begin {center}
    \caption{\textbf{Chi-Square Test for Level of Education and Cyber-hygiene behaviour}}
      \label{tab: table7}
       \begin{tabular}{l|c|c|r}
        \textbf{Quantity} & \textbf{Value} & \textbf{df} & \textbf{ Asym. Sig (2-sided)}\\
         \hline
          Pearson Chi-Square & 5.012 & 4 & 0.285\\
          Likelihood Ratio & 5.488& 4 & 0.241\\
          Linear-by-Linear Association & 0.135 & 1 & 0.714\\
         N of Valid Cases & 145 \\
      \end{tabular}
  \end {center}
\end {table}

\subsection{\textbf{Percentage Representation of Good Knowledge and Good Behaviour}}
Tables \ref{tab: table8} and \ref{tab: table9} show the descriptive statistics of good knowledge scores and good behaviour score respectively. This was done to find out if the data is normally distributed, that is, if the median is the same as the mean. There are other tests used for normality but are not included here. Both tables show that data are normally distributed. That is it failed the tests. So, the median was used as the cut off point for the categorisation of the respondents into those with poor and good knowledge, and poor and good behaviour. Those who had a score of 10 and above had good knowledge, while those with lower than 10 score were grouped as having poor knowledge. The median for cyber-hygiene behaviour score is 6, therefore, those who had a score of 6 and above had good behaviour, while those with lower than 6 score were grouped as having poor behaviour.
As shown in Table \ref{tab: table10}, the number of the respondents with good knowledge is 78, representing 53.79\%, while 75 have good behaviour, that is 71.72\%. This result shows that a reasonable number of the respondents, almost half, do neither have good cyber-hygiene knowledge nor good cyber-hygiene behaviour. \\

\begin {table}[h!]
  \begin {center}
    \caption{\textbf{Descriptive for cyber-hygiene knowledge score}}
      \label{tab: table8}
       \begin{tabular}{l|c|r}
  \textbf{Total cyber-hygiene knowledge score } & \textbf{ Statistic } & \textbf{Std. Error} \\
         \hline
          Mean & 9.39 & 0.123\\
          95\% confidence    upper range & 9.15\\
          Interval of mean   lower range & 9.64\\
          5\% Trimmed Mean & 9.49\\
          Median & 10.00\\
          Variance & 2.20\\                                         Std. Deviation & 1.48\\
         Minimum & 4.00\\
         Maximum & 12.00\\
         Range & 8.00\\
        Interquartile Range & 2.00\\
        Skewness & -0 .87 & 0.201\\
        Kurtosis & 0.81 & 0.400\\
      \end{tabular}
  \end {center}
\end {table}

\begin {table}[h!]
  \begin {center}
    \caption{\textbf{Descriptive for Cyber-hygiene behaviour score}}
      \label{tab: table9}
       \begin{tabular}{l|c|r}
  \textbf{ Total Cyber-hygiene behaviour score } & \textbf{ Statistic } & \textbf{Std. Error} \\
         \hline
          Mean & 5.84 & 0.248\\
          95\% confidence    upper range & 5.35\\
          Interval of mean   lower range & 6.33\\
          5\% Trimmed Mean & 5.78\\
          Median & 6.00\\
          Variance & 8.93\\                                     
          Std. Deviation & 2.99\\
         Minimum & 0.00\\
         Maximum & 13.00\\
         Range & 13.00\\
        Interquartile Range & 5.00\\
        Skewness & 0 .27 & 0.201\\
        Kurtosis & -0.78 & 0.400\\
      \end{tabular}
  \end {center}
\end {table}

\begin {table}[h!]
  \begin {center}
   \caption{\textbf{Percentage Representation of Good Knowledge and Good Behaviour }}
      \label{tab: table10}
       \begin{tabular}{l|c|c|r}
        \textbf{Descriptive variable} & \textbf{Total Sample} & \textbf{Good Knowledge} & \textbf{Good Behaviour}\\
        
$\hspace{1cm}$ & $N=145(100)\%$ & $N=75(51.72)\%$ & $N=78(53.79)\%$\\
         \hline
          \multirow {2} {*}
            \underline{Age}\\
            
          15-24yrs & 59(40.7) & 31(39.7) & 27(36.0)\\
          25-34yrs & 37(25.5) & 19(24.4) & 21(28.0)\\
          34-44yrs & 29(20.0) & 15(19.2) &17(22.7)\\
         45-54yrs & 16(11.0) & 9(11.5) & 9(12.0)\\
        Above 54yrs & 4(2.8) & 4(5.1) & 1(1.3)\\
        \hline
        Educational Level\\
        Diploma & 1(0.7) & 0(0.0) & 0(0.0) \\
        Undergraduate & 64(44.1) & 32(44.1) & 30(40.0)\\
        Graduate & 31(21.4) & 19(24.4) & 21(28.0)\\
        Masters & 29(20.0) & 15(19.2) & 14(18.7)\\
        Doctorate & 20(13.8) & 12(15.4) & 10(13.3)\\
      \end{tabular}
  \end {center}
\end {table}

\section{\textbf{Conclusion and Recommendation}}
\label{Section: Conclusion}
The primary target of our study was to determine the effect of age and educational level on the cyber-hygiene knowledge of employees and students of University of Nigeria, Nsukka, through an online pilot study. Thus, the detailed findings of the research were presented in the result and discussion section of our paper. From the results we have, it was found that the two variables: age and level of education do not have significant effect on the cyber-hygiene knowledge and behaviour of students and employees of University of Nigeria, Nsukka. It was also discovered that a reasonable proportion of the participants of this survey have poor cyber-hygiene knowledge and behaviour, which suggests the need for proper awareness and training. \\

The result of our findings as regards the effect of age on cyber-hygiene correlates with the result in obtained in \cite{cain2018exploratory}, which also found that age has no statistical significance with cyber-hygiene.
Based on the findings from this research, it can be recommended that well focused training on cyber-hygiene best practices should be organised for both students and employees of the tertiary institutions. Also, the internet security units of tertiary institutions should put in place adequate security infrastructures to safeguard the institution’s information and devices against cyber-attacks.\\

\section{\textbf{Future Work}}
\label{Section: Future Work}
In future study, the effect of other factors such as gender, level of exposure, area of discipline that might impact on the cyber-hygiene will be investigated. One major limitation of this study was poor responses occasioned by closed down of institutions because of Coronavirus disease 2019 (COVID-19) pandemic. \\

\bibliographystyle{IEEEtran}
\bibliography{Bibliography.bib}
\end{document}